\documentclass[twocolumn,prl,showpacs]{revtex4}

\usepackage{amsmath}
\usepackage{graphics}
\usepackage{epsfig}

\begin{document}

\title{Scattering Length Scaling Laws for Ultracold Three-Body Collisions} 
\author{J. P. D'Incao and B. D. Esry}
\affiliation{Department of Physics, Kansas State University,
Manhattan, Kansas 66506} 

\begin{abstract}
We present a simple picture that provides the energy and scattering
length dependence for all inelastic three-body collision rates in the ultracold
regime for three-body systems with short range two-body interactions.  In
particular, we present the scaling laws for 
vibrational relaxation, three-body recombination, and collision-induced
dissociation for systems that support $s$-wave two-body collisions.  These
systems include three identical bosons (BBB), two identical bosons (BBB$'$),
and two identical fermions (FFF$'$). Our approach reproduces all
previous results, predicts several others, and gives the general form
of the scaling laws in all cases.
\end{abstract} 
\maketitle

The use of external magnetic fields to control the atomic interactions 
in trapped ultracold quantum gases has proven to be an extraordinary tool to
explore different quantum regimes. At low temperatures, only the two-body
$s$-wave scattering length $a$ is needed to characterize the atomic
interactions and can assume practically any value from $-\infty$ to
$+\infty$ by tuning a magnetic field near a diatomic Feshbach
resonance. This tunability has been used, for instance, to convert
fermionic atoms into weakly bound bosonic molecules which, in some
cases, were remarkably long lived
\cite{Jin}. In the quantum degenerate
regime, this system provides a unique opportunity to explore the
crossover between a Bose-Einstein condensate (BEC) of molecules and
the BEC of atomic Cooper pairs, the BEC-BCS
(Bardeen-Cooper-Schrieffer) transition.  

Since the molecular lifetimes are influenced by three-body
processes, these experiments underscore the importance of knowing the
dependence of ultracold three-body collision rates on $a$.
In particular, vibrational relaxation, X+(XX)$^*$$\rightarrow$X+XX,
releases enough kinetic energy to free the collision products from
typical traps, leading to molecular loss. Collision induced
dissociation, X+XX$\rightarrow$X+X+X, can also contribute to molecular
loss for large $a$. The other possible three-body process, three-body
recombination X+X+X$\rightarrow$X+XX, contributes to atomic losses. 
In an experiment, however, the dominant process is determined in part
by the threshold and scattering length scaling laws.

Threshold laws, which give the energy dependence for small collision
energies, dictate the partial wave that dominates for each process.
The threshold law for elastic two-body collisions, for instance, gives
an $s$-wave cross section that is constant at threshold while the
$l$-th partial wave is suppressed by a factor of $E^{2l}$. Combined
with permutation symmetry requirements, the threshold law thus leads
to the familiar conclusion that the collision cross section for two
indistinguishable fermions vanishes at low energies. Three-body
threshold laws similarly depend on the number and kind of identical
particles for each partial wave $J^\pi$, where $J$ is the total
orbital angular momentum and $\pi$ is the parity~\cite{Threshold}.  

While general results exist for three-body threshold laws, no
similarly general scattering length scaling laws have yet been  
obtained. Specific cases are known: the recombination rate for 
B+B+B collisions, where B is a bosonic atom, 
scales roughly as $a^4$ for 0$^+$ and all
$a$~\cite{NielsenEsry01,Braaten01}, and as $a^8$ for $2^{+}$ and
$a$$>$$0$ \cite{Dincao01}. It has also been shown for $a$$>$$0$ that the
relaxation rate for B+BB collisions is linear in $a$ \cite{Braaten02},
but is $a^{-3.33}$ for F+FF$'$ collisions~\cite{Petrov}, where F and
F$'$ are distinguishable fermionic atoms and $a$ is the F+F$'$
$s$-wave scattering length. This scaling law is qualitatively
different from the other known scaling laws and begs the question: how
do the three-body processes for other systems scale with $a$? 

In this Letter we present a simple physical picture within which both 
the energy and scattering length dependence of all ultracold
three-body collision rates (for short-range two-body interactions) can
be derived and understood. A simple picture emerges because all
three-body systems can be represented by one of four prototype systems
(two each for $a$$>$$0$ and $a$$<$$0$). In all cases, though, the rate
limiting step is tunneling through a potential barrier in the initial
channel. The barrier is determined here from effective three-body
potentials obtained from the adiabatic hyperspherical representation
\cite{Esry02}. A simple WKB approximation to the tunneling probability
is then sufficient to give both the energy and scattering length 
dependence. Besides their practical use, the scaling laws make
explicit the pervasive influence of Efimov physics~\cite{Efimov} on
ultracold three-body collisions.  We restrict our discussion to
systems with equal masses and to symmetries that support $s$-wave
two-body collisions, representing most cases of experimental
interest.  

In the adiabatic hyperspherical representation, the three-body
effective potentials and couplings are determined from the
adiabatic equation \cite{Esry02},
\begin{equation}
H_{\rm ad}(R,\Omega)\Phi_\nu(R;\Omega)=U_\nu(R) \Phi_\nu(R;\Omega),
\label{adeq}
\end{equation}
where $\Omega$ denotes all hyperangles and $R$ is the
hyperradius that, roughly speaking, gives the overall size of the
system. The adiabatic Hamiltonian $H_{\rm ad}$ includes the hyperangular
kinetic energy as well as all interactions. By expanding the total
wave function on the adiabatic basis $\Phi_\nu$, the Schr{\"o}dinger
equation is reduced to (atomic units will be used unless otherwise
noted):  
\begin{eqnarray} 
\left[-\frac{1}{2\mu}\frac{d^2}{dR^2}+W_{\nu}\right]F_{\nu}
+\sum_{\nu'\neq\nu} V_{\nu\nu'} F_{\nu'}=E F_\nu.  
\label{radeq}
\end{eqnarray}
In this expression, $\mu$ is the three-body reduced mass, $E$ is the
total energy, $F_{\nu}$ is the hyperradial wave function, 
$V_{\nu\nu'}$ is the nonadiabatic coupling responsible for inelastic
transitions, and $W_{\nu}$ is the effective potential. 

For short range interactions, the asymptotic behavior of $W_{\nu}$ can
be derived analytically~\cite{NielsenReview}. In this limit, the
molecular channels, which represent atom-molecule scattering, 
and the three-body continuum channels, which represent collisions of
three free particles, are given by  
\begin{equation}  
W_{\nu}={E}_{vl'}\!+\!\frac{l(l\!+\!1)}{2\mu R^2}
~~\mbox{and}~~  W_{\nu}=\frac{\lambda(\lambda\!+\!4)+\!15/4}{2\mu R^2},
\label{bcch}
\end{equation}  
respectively.
The molecular bound state energy ${E}_{vl'}$ is labeled by the
ro-vibrational quantum numbers $v$ and $l'$;  $l$ is the relative
angular momentum between the molecule and the atom; and $\lambda$ is a
positive integer that labels the eigenstates of the hyperangular
kinetic energy. 

When the $s$-wave scattering length $|a|$ is large, however, an
intriguing phenomenon known as the Efimov effect occurs \cite{Efimov},
modifying the behavior of $W_\nu$. These modifications must be
considered in order to properly predict the dependence of the
three-body rates on $a$
\cite{NielsenEsry01,Braaten01,Dincao01,Braaten02}. In the limit
$|a|$$\gg$$ r_0$, where $r_0$ is the characteristic size of the 
two-body potential, Eq.~(\ref{bcch}) applies only for $R$$\gg$$|a|$.     
For $r_{0}$$\ll$$ R$$ \ll $$|a|$, the potentials are proportional
to $R^{-2}$, but can now be attractive as well as repulsive. 
Strictly speaking, the term ``Efimov effect'' applies only to the 
emergence of an infinity of three-body bound states for
$|a|$$\rightarrow$$\infty$.  We will instead use the term ``Efimov
physics'' to indicate the qualitative change in behavior exhibited by
any system whenever at least two of the three possible $s$-wave
scattering lengths is large.

Based on the modifications due to Efimov physics, we can classify all
three-body systems into one of two categories: those with an
attractive dipole potential for $r_{0}$$\ll$$ R$$ \ll$$ |a|$ and those
without.  For equal mass systems, only $0^{+}$ bosonic systems fall
into the first category. In these systems, the attractive potential
appears in the highest vibrationally excited $s$-wave (or weakly
bound) molecular channel ($l'$$=$$0$) for $a$$>$$0$, and in the lowest
continuum channel for $a$$<$$0$. The effective potentials for all higher
channels are repulsive, but the coefficients differ from those shown 
in Eq.(\ref{bcch}) due to Efimov physics. These effective potentials
are conveniently parametrized by, 
\begin{eqnarray}
W_{\nu}(R)=\frac{-s^{2}_{0}-\frac{1}{4}}{2\mu R^2}
\hspace{0.25cm}\mbox{and}\hspace{0.25cm}
W_{\nu}(R)=\frac{s^{2}_{\nu}-\frac{1}{4}}{2\mu R^2}.
\label{bcchefbosonic} 
\end{eqnarray}
The constants $s_0$ and $s_{\nu}$ depend on the number of resonant
pairs as well as the number of identical particles. For all other
cases, including $J$$>$$0$, the effective potentials are repulsive in
the weakly bound and continuum channels, for $a$$>$$0$ and $a$$<$$0$.
Like the $0^+$ bosonic case, these potentials are parametrized by constants
$p_0$ and $p_{\nu}$: 
\begin{eqnarray}
W_{\nu}(R)=\frac{p^{2}_{0}-\frac{1}{4}}{2\mu R^2}
\hspace{0.25cm}\mbox{and}\hspace{0.25cm}
W_{\nu}(R)=\frac{p^{2}_{\nu}-\frac{1}{4}}{2\mu R^2}.
\label{bccheffermionic} 
\end{eqnarray}
In all cases, deeply bound molecular channels are essentially
independent of $a$.

The constants $s_0$, $s_{\nu}$, $p_0$, and $p_{\nu}$ 
can be obtained analytically~\cite{Efimov}, and 
numerical values are shown in Table~\ref{TabEf} 
for the two partial waves that
dominate relaxation and recombination near threshold.
Table~\ref{TabEf} also shows the minimum $l$ and $\lambda$
[see Eq.~(\ref{bcch})] allowed by permutation symmetry that, in turn, determine
the dominant partial wave for the three-body rates near
threshold~\cite{Threshold}. 
\begin{table}
\begin{tabular}{c|ccc|cc} \hline\hline
 & $J^{\pi}$ & $l_{\rm min}$ & ~$\lambda_{\rm min}$~ & $s_{0}$ ($p_0$) & $s_{\nu}$ ($p_\nu$) \\ \hline
BBB &  0$^+$ & 0 & 0 & 1.0062378 & 4.4652946  \\
    &  1$^-$ & 1& 3 & 2.8637994 & 6.4622044 \\
    &  2$^+$ & 2 & 2 & 2.8233419 & 5.5082494 \\ \hline
BBB$'$& 0$^+$ & 0 & 0 & 0.4136973 & 3.4509891  \\
      & 1$^-$ & 1 & 1 & 2.2787413 & 3.6413035 \\ \hline
FFF$'$& 0$^+$ & 0 & 2 & 2.1662220 & 5.1273521  \\
      & 1$^-$ & 1 & 1 & 1.7727243 & 4.3582493 \\
      & 2$^+$ & 2 & 2 & 3.1049769 & 4.7954054 \\
\hline\hline
\end{tabular}
\caption{Coefficients of the potentials in
Eqs.~(\ref{bcch})--(\ref{bccheffermionic}). 
Except for $0^+$ bosons, the constants correspond 
to $p_0$ and $p_\nu$.}
\label{TabEf}
\end{table}

\begin{figure}[htbp]
\includegraphics[width=2.8in,angle=0,clip=true]{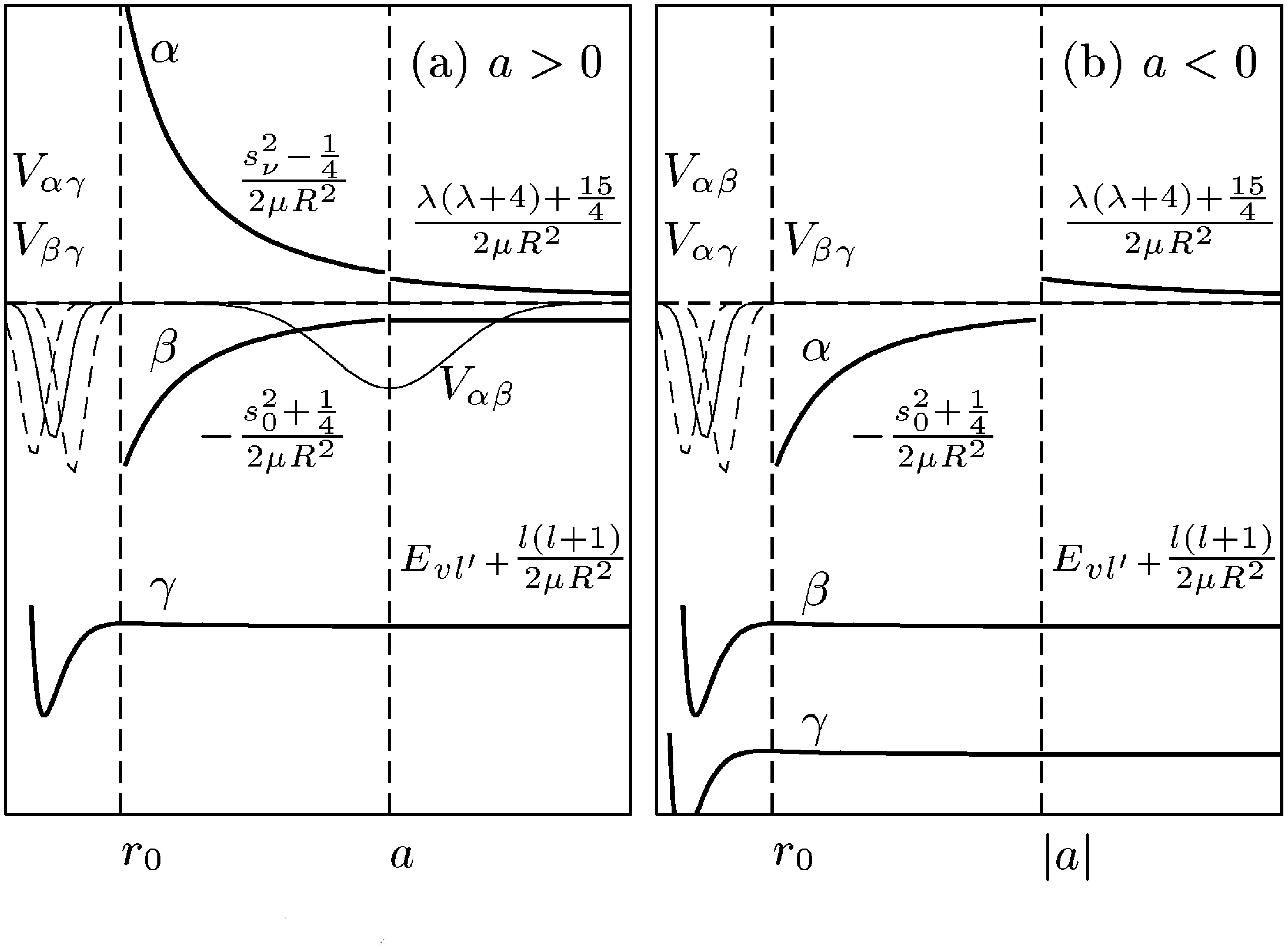}
\vskip -0.1 in
\includegraphics[width=2.8in,angle=0,clip=true]{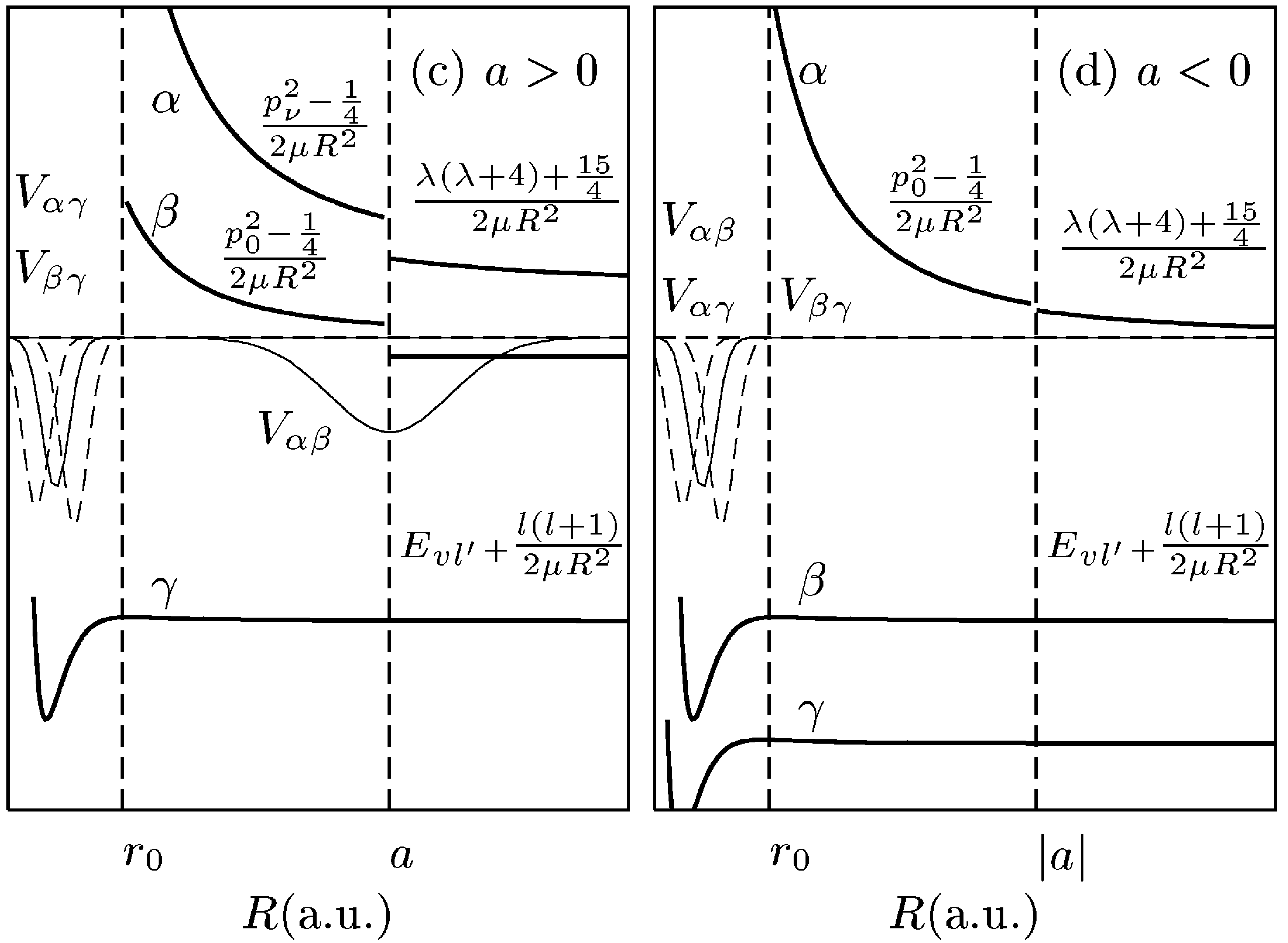}
\caption{Schematic plot of the three-body effective potentials
  for $0^{+}$ bosonic systems for (a) $a$$>$$0$ and (b)
  $a$$<$$0$, and for the others systems, for (c) $a$$>$$0$ and (d) $a$$<$$0$.
  \label{Fig1}} 
\end{figure}

The relevant effective potentials and couplings are sketched in
Figs.~\ref{Fig1}(a) and \ref{Fig1}(b) for $0^+$ bosonic systems and in
Figs.~\ref{Fig1}(c) and \ref{Fig1}(d) for all other cases.  For
$r_0$$<$$R$$<$$|a|$, the potential curves are given by
Eqs.~(\ref{bcchefbosonic}) and (\ref{bccheffermionic}); for
$R$$>$$|a|$, the curves are those for finite $|a|$ in
Eq.~(\ref{bcch}). The lowest continuum channel, labeled ``$\alpha$'', 
is the initial channel for recombination.  Channel ``$\beta$'' is the
weakly bound molecular channel and is the initial channel for
relaxation. Channel ``$\gamma$'' is a deeply bound molecular channel
in all cases. The nonadiabatic couplings are sketched in
Fig.~\ref{Fig1} and indicate the regions where inelastic transitions
are most likely. From our numerical calculations and on physical
grounds, we believe that Fig.~\ref{Fig1}
represent all three-body systems near threshold.  

Knowing the dependence of the potentials and couplings on $R$ and $a$ 
enables us to derive the scaling laws from the definitions of the rates,
\begin{equation}
V_{\rm rel} \propto {|T_{fi}|^2}/{k} \quad\mbox{and}\quad
K_{3} \propto {|T_{fi}|^2}/{k^4},
\label{RateDefs}
\end{equation}
in terms of the transition probabilities $|T_{fi}|^2$.
Only the wave vector important for the threshold
law has been included --- $k^2$=$2\mu (E-E_{vl'})$ for $V_{\rm rel}$ and
$k^2$=$2\mu E$ for $K_3$. The transitions proceed via tunneling in the
initial potential to the $R$ where the coupling peaks.  The
transition probability can thus be approximated by the WKB tunneling
probability (including the Langer correction~\cite{Langer}), 
\begin{eqnarray}
P^{(\nu)}_{x\rightarrow y} \! \approx
\!\exp\!\left[-2\!\!\int_{y}^{x}\!\!\sqrt{2\mu\!\left(\!W_{\nu}(R)+
\frac{{1}/{4}}{2\mu R^2}-E\!\right)}dR\right].
\label{TransProb}
\end{eqnarray}
The Langer correction is crucial for obtaining the correct
scaling with $E$ and $a$, and introduces ``tunneling'' even when there
is no barrier in $W_{\nu}$.  At ultracold temperatures, the classical
turning point is much larger than $|a|$.

In all cases, the
couplings peak at $R$$\approx $$|a|$, $R$$\approx $$r_0$, or both.
Relaxation for $a$$>$$0$, for instance, only occurs at small
distances, $R$$\approx $$r_{0}$, where the coupling peaks
[$V_{\beta\gamma}$ in Figs.~\ref{Fig1}(a) and \ref{Fig1}(c)]. It
follows that the tunneling probability in the initial channel $\beta$
must be evaluated by integrating Eq.~(\ref{TransProb}) from the
classical turning point $r_c$ to $r_{0}$. This range, however, spans
both kinds of potentials so that the tunneling in the two regions 
must be included,
\begin{eqnarray}
|T_{\gamma\beta}|^2 \simeq 
P^{(\beta)}_{r_{c}\rightarrow a}P^{(\beta)}_{a\rightarrow r_{0}}.
\label{TVrelapos}
\end{eqnarray}
Similarly, the transition probability for recombination must also include
tunneling in two different regions between $r_{c}$ and $r_{0}$.  Recombination for $a$$>$$0$
can occur at two distances, though --- $R$$\,\approx\, $$a$ and 
$R$$\,\approx\, $$r_{0}$ --- since the coupling peaks both places,
leading to    
\begin{eqnarray}
|T_{\beta\alpha}|^2 \simeq 
P^{(\alpha)}_{r_{c}\rightarrow a}+
P^{(\alpha)}_{r_{c}\rightarrow a}P^{(\beta)}_{a\rightarrow r_{0}}+ 
P^{(\alpha)}_{r_{c}\rightarrow a}P^{(\alpha)}_{a\rightarrow r_{0}},
\label{TK3D3apos}
\end{eqnarray}
where each term corresponds to a different reaction pathway. Although
not indicated here, these paths can interfere and will be discussed
below. For $a$$<$$0$, all couplings peak at small distances
$R$$\,\approx\,$$ r_{0}$.  Applying the arguments above, relaxation
and recombination are, respectively, 
\begin{equation}
|T_{\gamma\beta}|^2 \simeq P^{(\beta)}_{r_{c}\rightarrow r_{0}} 
\quad\mbox{and}\quad
|T_{\beta\alpha}|^2 \simeq 
P^{(\alpha)}_{r_{c}\rightarrow |a|}
P^{(\alpha)}_{|a|\rightarrow r_{0}}.
\label{TK3D3aneg} 
\end{equation}

We can now determine both the threshold laws and the scattering length
scaling.  Since $W_\nu$ has the same form in the region $R$$>$$|a|$ in 
all cases [Eq.~(\ref{bcch})], the tunneling probabilities in this
region are also the same,  
\begin{eqnarray}
P^{(\alpha)}_{r_{c}\rightarrow |a|} \:{\propto}\:
(ka)^{2\lambda+4} ~~\mbox{and}~~
P^{(\beta)}_{r_{c}\rightarrow |a|} \:{\propto}\: (ka)^{2l+1}.
\label{TPbeta} 
\end{eqnarray}   
Equation (\ref{TPbeta}) completely determines the 
threshold laws for each process~\cite{Threshold}. 
The scaling with $a$, however, will be strongly modified by the tunneling probability in the region 
$r_{0}$$<$$R$$<$$|a|$ due to Efimov physics.


For $0^+$ bosons with $a$$>$$0$, 
the $\beta$-channel potential is attractive in the region $r_0$$<$$R$$<$$a$.
Consequently, the probability $P_{a\rightarrow r_0}^{(\beta)}$
[Eqs.~(\ref{TVrelapos}) and (\ref{TK3D3apos})] is not a tunneling
probability, but rather a transmission probability that can be
determined from general arguments based on the known solutions for
Eq.~(\ref{bcchefbosonic})~\cite{Braaten02,Efimov}. This analysis gives
\begin{eqnarray}
&V_{\rm rel}\propto
\left[\frac{\sinh(2\eta)}{\sin^2[s_{0}\ln({a}/{r_{0}})+\Phi]+\sinh^2(\eta)}\right]a,& 
\label{BosonicVRapos}
\end{eqnarray}
since $l$$=$$0$ gives the leading contribution in
Eq. (\ref{TPbeta}). The constant $\eta$ is related to the probability
for transitions at small distances~\cite{Braaten02} and $\Phi$ is an
unknown small-$R$ phase. Equation~(\ref{BosonicVRapos}) was deduced in 
Ref.~\cite{Braaten02}, generalizing the result presented in
Ref.~\cite{Efimov} by letting the transition probability at small
distances be less than unity. In both cases, however, the linear
dependence on $a$ was obtained by dimensional and physical arguments,
while it follows directly from Eq.~(\ref{TPbeta}) in the present
analysis.  

For 0$^+$ bosons with $a$$<$$0$, the initial state for relaxation is a  
deeply bound vibrational state independent of $a$. The transmission
probability --- and thus the relaxation rate --- does not depend on
$a$ either, leading to  
\begin{eqnarray}
V_{\rm rel}\propto
A_{\eta}k^{2l}{r_{0}^{2l+1}}. 
\label{vibrelanegA}
\end{eqnarray}
The constant $A_{\eta}$ will generically represent small-$R$ phy\-sics
that can give resonances due to a three-body Feshbach resonance, but
is otherwise independent of $a$.


For all cases other than $0^+$ bosons, the initial channel for
relaxation is never attractive in the region $r_0$$<$$R$$<$$|a|$
[Figs.~\ref{Fig1}(c) and \ref{Fig1}(d)] so that relaxation proceeds by
tunneling only. For $a$$>$$0$, the relaxation rate is 
\begin{eqnarray}
&V_{\rm rel}\propto 
A_\eta k^{2l}\left(\frac{r_{0}}{a}\right)^{2p_{0}}{a^{2l+1}}.&
\label{FermionicVRapos}
\end{eqnarray}
The rate thus decreases with $a$ whenever $2l$$+$$1$$<$$2p_{0}$,
yielding longer molecular lifetimes for larger $a$.  For the specific
case of a weakly bound FF$'$ $s$-wave molecule, $p_{0}$$=$$2.1662220$
so that $V_{\rm rel}$$\,\propto\,$$a^{-3.332444}$. This result agrees
with the recent prediction of Petrov {\it et al.}~\cite{Petrov} and is
consistent with experiments \cite{Jin}. Like the 0$^+$ boson systems, 
relaxation does not depend on $a$ for $a$$<$$0$, 
\begin{eqnarray}
V_{\rm rel}\propto 
A_{\eta}k^{2l}{r_{0}^{2l+1}}. 
\label{vibrelanegB}
\end{eqnarray}


The present analysis applies equally well to recombination. For
instance, recombination of $0^+$ bosons with $a$$>$$0$ is determined
from Eq.~(\ref{TK3D3apos}), which includes three different
recombination paths. Only the interference between the first two terms
in Eq.~(\ref{TK3D3apos}) will be included here, since we expect the
third term will be suppressed. In fact, this interference is well
known \cite{NielsenEsry01,Braaten01}, and the present analysis
reproduces the known expression plus a modification:   
\begin{eqnarray}
&K_{3} \propto \left[
\sin^{2}\left[s_{0}\ln({a}/{r_{0}})+\Phi\right]+
A_{\eta}\left(\frac{r_{0}}{a}\right)^{2s_{\nu}}\right]a^{4}.&
\label{BosonicK3pos}
\end{eqnarray}
The first term in Eq.~(\ref{BosonicK3pos}) is the usual result, while
the second is due to recombination at small $R$. Because the small-$R$
coupling lies so far into the classically forbidden region, we expect
$A_{\eta}$ to be small. For $a$$<$$0$, the present analysis yields the
same expression found in Ref.~\cite{Braaten01}, 
\begin{eqnarray} 
&K_{3}\propto
\left[\frac{\sinh(2\eta)}{\sin^2\left[s_0\ln({|a|}/{r_{0}})+\Phi\right]+\sinh^2(\eta)}\right]
a^{4}.&
\label{BosonicK3neg}
\end{eqnarray}
The recombination rate for all other cases
can be determined, giving for $a$$>$$0$ and $a$$<$$0$, respectively,   
\begin{eqnarray}
&K_{3}\propto
k^{2\lambda}\left[1
+A_{\eta}\left(\frac{r_{0}}{a}\right)^{2p_{0}}
+B_{\eta}\left(\frac{r_{0}}{a}\right)^{2p_{\nu}}\right]a^{2\lambda+4},&
\label{OthersK3pos}\\
&K_{3}\propto
k^{2\lambda}\left(\frac{r_{0}}{|a|}\right)^{2p_{0}}|a|^{2\lambda+4},& 
\label{K3otheraneg}
\end{eqnarray}
\noindent 
predicting an asymmetry in $K_3$ for $a$$<$$0$ and $a$$>$$0$. The
constants $A_{\eta}$ and $B_{\eta}$ are expected to be small.  

\begin{table}
\begin{tabular}{c|c|ccc|ccc}\hline\hline
  &  & \multicolumn{3}{c}{$V_{\rm rel}$} &  \multicolumn{3}{c}{$K_{3}$ $(D_{3})$} \\ 
  &$J^{\pi}$ & $E$ & $a>0$ & $a<0$ & $E$ & $a>0$ & $a<0$   \\ \hline
BBB &$0^+$ & {\bf const} & \boldmath{$a$} & {\bf const} &  {\bf const}
  \boldmath{$(k^{4})$} & \boldmath{$a^{4}$}  & \boldmath{$|a|^{4}$}  \\ 
    &$1^-$ & $k^{2}$ & $a^{-2.728}$ & const & $k^{6}$ $(k^{10})$ & $a^{10}$ & $|a|^{4.272}$  \\
    &$2^+$ & $k^{4}$ & $a^{-0.647}$ & const & $k^{4}$ $(k^{8})$ & $a^{8}$  & $|a|^{2.353}$  \\ \hline
BBB$'$ & $0^+$ &{\bf const} &\boldmath{$a$} & {\bf const} & {\bf const} \boldmath{$(k^{4})$} & \boldmath{$a^{4}$} & \boldmath{$|a|^{4}$}   \\
       & $1^-$ & $k^{2}$ & $a^{-1.558}$ &const &  $k^{2}$ $(k^{6})$ & $a^{6}$  & $|a|^{1.443}$   \\ \hline
FFF$'$ &$0^+$  & {\bf const} & \boldmath{$a^{-3.332}$} & {\bf const} & $k^{4}$ $(k^{8})$ & $a^{8}$  & $|a|^{3.668}$\\
       &$1^-$  & $k^{2}$ & $a^{-0.546}$ & const & \boldmath{$k^{2}$ $(k^{6})$} & \boldmath{$a^{6}$}  & \boldmath{$|a|^{2.455}$}\\
       &$2^+$      & $k^{4}$ & $a^{-1.210}$ & const & $k^{4}$ $(k^{8})$ & $a^{8}$  & $|a|^{1.790}$\\
\hline\hline
\end{tabular}
\caption{Threshold and scattering length scaling laws for three-body
  rates. Boldface indicates dominant contributions.\label{TabRates}}
\end{table}

Our numerical results for B+BB and F+FF$'$ collisions confirm
Eqs.~(\ref{BosonicVRapos})--(\ref{vibrelanegB}). In fact, in our model  
for F+FF$'$ the contributions for both $s$- and $p$-wave final states
scale as $a^{-3.33}$, emphasizing that the scaling only depends on
the initial state. Equations~(\ref{BosonicK3pos})--(\ref{OthersK3pos})
have been verified numerically and their limitations
studied~\cite{Dincao01}. 

Table~\ref{TabRates} summarizes the scaling laws, showing only
the main power-law behavior of each rate. The two dominant partial
waves, determined by their energy dependence, are shown for each
process. For completeness, we include the scaling laws for the
dissociation rate $D_3$ (for which $k^2$=${2\mu E}$). It is
interesting to note that for F+FF$'$ collisions where the relaxation
rate decreases with $a$, the dissociation rate grows (see
Table~\ref{TabRates}) and will eventually become the dominant
mechanism of molecular loss. The table also indicates the competition
from the next leading term for finite temperatures.  For fermion
relaxation, this term is comparatively more important than for boson
relaxation. 

In this Letter, we have deduced the scaling laws for ultracold
three-body collision rates for all symmetries.  We have used a simple
and intuitive approach that describes all three-body collision
processes in the same framework. For relaxation and recombination,
the scaling laws depend only on the initial state.  Any dependence on
the final state is expected to enter via the coupling terms, and
should at most be weakly dependent on energy and scattering length.  The
present results apply only in the threshold regime, i.e. when the
collision energy is the smallest energy in the system \cite{Dincao01}. We 
have also shown the remarkable influence of Efimov physics --- even
for systems without bosons.

This work was supported by the National Science Foundation and by the
Research Corporation.

\end{document}